# Optical Characterization of Single Mode Components by new Medianfield-method

**Th. Windel, U. H. P. Fischer (member IEEE)**
*Harz University of Applied Sciences, Friedrichstr. 57, 38855 Wernigerode, Germany,*
*E-mail:ufischerhirchert@hs-harz.de*

**Abstract:** In this paper a new method for spotsize-measurement for singlemode optical components is presented. Based from the classical farfield-method where the measurements are made circular, the mounting of the used rotary stages and the long measurement time are great disadvantages. In this paper a new planar method is described which overcomes these problems. Based on the measurement of a singlemode fiber in accordance with ITU Recommendation G.652 the efficiency is demonstrated and discussed.

## 1. Introduction

The demand for high-speed digital communication such as data, video, and the broadband Internet increases, the required throughput of the modules in communications systems[1] will also increase. Fast transmitter and receiver modules are basic elements of these systems, which should be able to transmit terabits/s of information via the fiber. Such technologies in turn rely strongly on advanced opto-electronically technologies, and the progress made in integrated optics. With rapidly increasing market demand for the use of these technologies and access to the information highway, the next challenge in optical communications is to provide high coupling efficiency in stable modules[2] at affordable prices.

In this paper we present a new method for spotsize-measurement for singlemode optical components. Based from the classical far field-method where the measurements are made circular, the mounting of the used rotary stages and the long measurement time are great disadvantages. In this paper a new planar method is described which overcomes these problems.

In the first section the conversion of the values of planar measurement to values of circular measurement is specified, while in the following experimental part a Labview program was developed to implement the planar measurement-method and convert the data for an existing 6-axes-motion controller. Based on the measurement of a singlemode fiber in accordance with ITU Recommendation G.652 the efficiency is demonstrated and discussed.

## 2. OEIC-fiber coupling

A comparison of the optical fields of a cleaved standard monomode fiber (SMF) and of InP integrated wave-guides shows great mode fields mismatch (fig. 1). This mismatch is the reason for the very low coupling efficiency[3.] of around 10 % (_10 dB) This low efficiency can be overcome by a better adaptation of the two optical mode fields with mode field transformers. Normally, optical lenses realized these transformers, which are integrated within the SMF-end face so called fiber taper lenses[4] (fig. 2) with a coupling efficiency of more than 50 % (_3 dB).

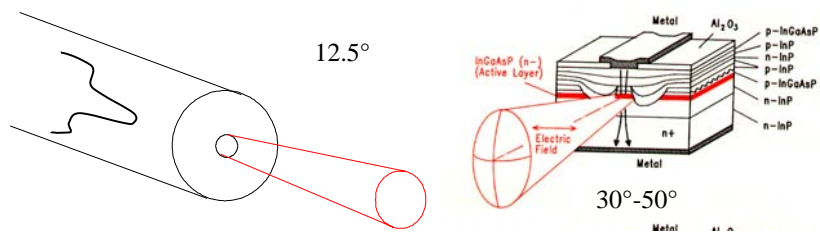
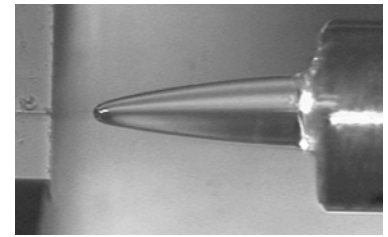

Fig. 1: Optical far field of SMF and laser diodes

Fig. 2: Tapered SMF fiber

To adapt the outline of the fiber taper the optical mode field of the laser diode must be characterized very exactly. In the following different methods for mode field acquisition are described and a new technique will be proposed.

## 3.  Mode field measurement

There are several methods to characterise the optical mode field of photonic devices[5, 6]. The most used method is the so-called far-field method[7]. Here the optical far field is measured by a scanning photodiode, which is rotated around the output side of optical waveguide. The measurement is performed in a long distance of several cm from the end face of the waveguide that can be realized without any additional optics and simple rotating stages. This method is also described in the EIA standard RS-455-47[8] and is used standardised for qualifying optical fibers. Here the resolution of the method is limited by the scanning resolution and additionally in the far field a smearing of the fine contour of the near field can be observed. The set-up of this method is relative simple but needs a large room around the DUT, because the detector photodiode needs several cm scanning space.

The second often-used method is using a microscope objective to enlarge the surface spot of the waveguide front to a camera system. This method is called in literature the near field technique[9]. The resolution of this method is strongly limited by the diffraction limit of the used light of the laser diode, which is fed through the tested waveguide. At 1550nm usually the resolution limit is in the region of 2µm in both lateral dimensions. For typical optical fields of laser diodes the resolution of this method is not sufficient enough to get a detailed insight of the optical field. Problems are additionally the lens aberrations, which broaden the focus on the camera screen and the DUT must be allocated in micrometer precision in front of the set-up that takes very long time and limits this method to laboratory use only

Both methods have the advantage of long measurement and positioning times for the operator, which constricts these methods for high output, automated industrial use.

New Medianfield-method

To overcome the disadvantages of the conventional methods, we developed a new set-up for scanning the optical field linearly without any additional optics or µm-precise adjustment of the DUT.

The measurement of the intensity takes place with the help of a photodiode in a 2-axis system, whereby the distance admits r of the level to the test object be must. See fig. 3. The distance r can be seized on use of a 3-axis system directly from the data of the traverse path. It does not need to be changed during the measurement.

In relation to the classical far field method a conversion of the measured intensities must be made, since the distance of the photodiode from the Z-axis no longer directly aligns the photodiode to the test object. It would have to be turned, in order to keep the effective receipt surface constant. Additionally the distance of the photodiode to the test object changes. Both factors affect thus the effective surface of the measurement-receiving device and must be considered by a distance coefficient.

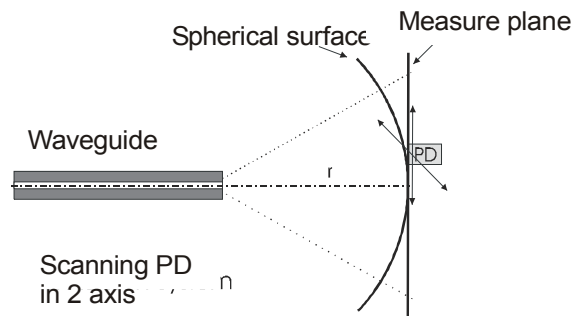

Fig 3: Medianfield measurement set-up.

The coefficient is determined uniquely for each measuring point and considered when later on measurements were performed again. The surface of the measurement receiving device (usually a photodiode) is assumed here as rectangular. The corner points can be regarded as vectors, for whose origin in the point of light withdrawal of the test object are appropriate, see fig. 4. The surface shown in the illustration corresponds thereby to the surface of the photodiode

By standardisation of the vectors on the length of the vector, which represents the center of the rectangle, one gets the effectively illuminated surface depicted schematically in fig. 4.

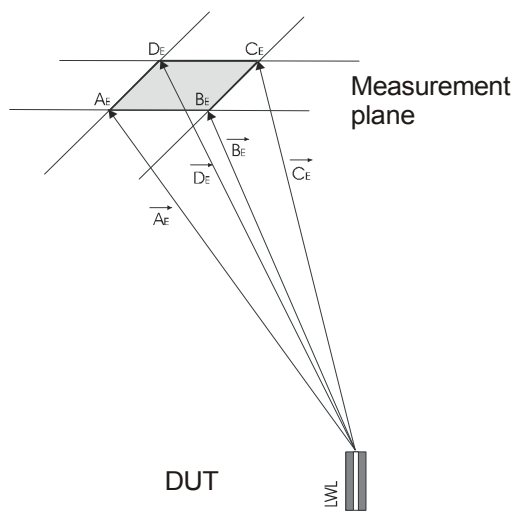
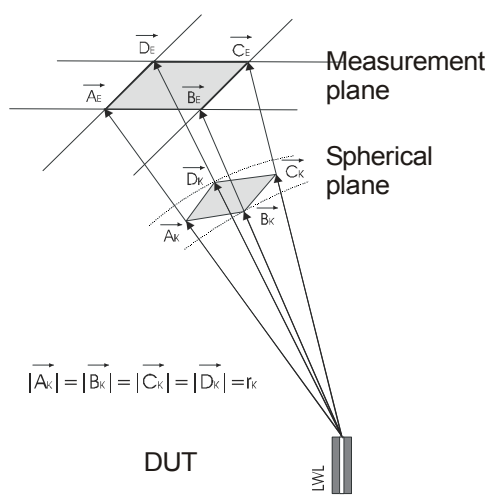

Fig. 4: Reception plane and DUT

Fig. 5: Difference of plane areas of flat measurement surface and spherical surface

The distance the photodiode reduces this surface from the Z-axis. It is also to be considered that the surface also must be rotated, which becomes an irregular square in many places in the measurement plane. The surface is determined according to the formula for the general square:

$$A = \frac{1}{2} * d_1 * d_2 * \sin \angle (d_1, d_2) \qquad (1)$$

d are the diagonals between the corner points of the effective measuring surface. With this the surface can be calculated by:

$$A_{EN} = \frac{1}{2} * \left| \overrightarrow{A_{EN} C_{EN}} \right| * \left| \overrightarrow{B_{EN} D_{EN}} \right| * \sin \angle \left( \overrightarrow{A_{EN} C_{EN}}, \overrightarrow{B_{EN} D_{EN}} \right) \qquad (2)$$

The difference of a flat plane area and the spherical one is depicted in fig. 5. For the determination of the surface on the spherical surface the vectors of the corner points of the measurement-receiving device are standardized on the radius of the ball. The radius corresponds thereby exactly to the distance of the measurement receiving device from the test object on the Z-axis. It stretches to an irregular square, whose surface can be determined as:

$$A_K = \frac{1}{2} * \left| \overrightarrow{A_K C_K} \right| * \left| \overrightarrow{B_K D_K} \right| * \sin \angle \left( \overrightarrow{A_K C_K}, \overrightarrow{B_K D_K} \right) \qquad (3)$$

For the determination of the distance coefficient both surfaces are set in relationship. This factor depends on the coordinates, where the measurement-receiving device is working in the measurement plane

$$\kappa(coordinates) = \frac{A_{EN}}{A_K} \qquad (4)$$

With this factor the intensities measured at the respective coordinates must be calibrated, in order to receive the real intensity values.

## 4. Automated acquisition with Labview

To the control of the used 6-axis-motion system (Physic Instruments (PI) F-604), which moves the photodiode device in the measurement plane, a Labview program was developed. It also accomplishes the determination of the distance coefficient. Here a virtual instrument (VI) was developed, which controls the 6-axis-motion system and performs the compilation of the measuring data. In fig. 6 the user front end is represented.

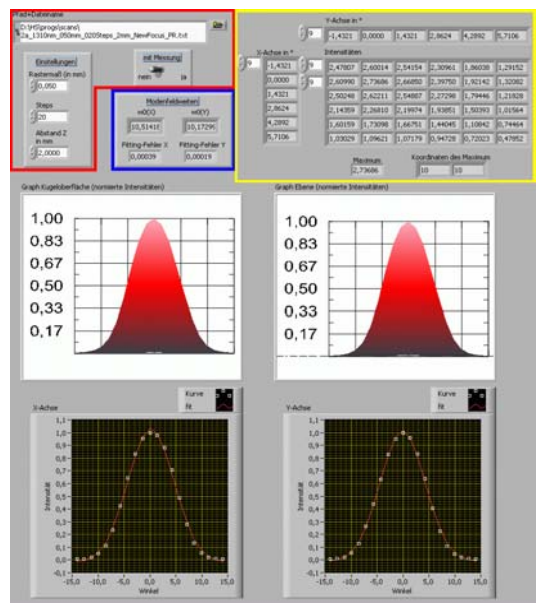

Fig. 6: LabVIEW VI user front end for automated field acquisition

In the left upper red bordered part the user is requested for the input of the file name for the storage of the measuring data. Further on the user is asked to type the incrementation steps size in µm and the number of measurements per line ("steps") into the marked fields. These data are needed for the automatic collection of the measuring data in combination of the drivers developed by PI. Within the yellow bordered range of the surface on the right upper side one can see the appropriate intensities converted with the distance coefficient. Further on the maximum intensity and their coordinates are indicated in the table. The range of the blue framework shows the optical field widths in x and y- direction, determined by the fitting process. In the lower figures the intensity field and the result of the fitting process are to be recognized.

## 5.  Measurement set-up

For the examination of the method a SMF manufactured after ITU-G.652 was measured. This reference measurement was realized with the following measurement set-up, shown in fig. 7:

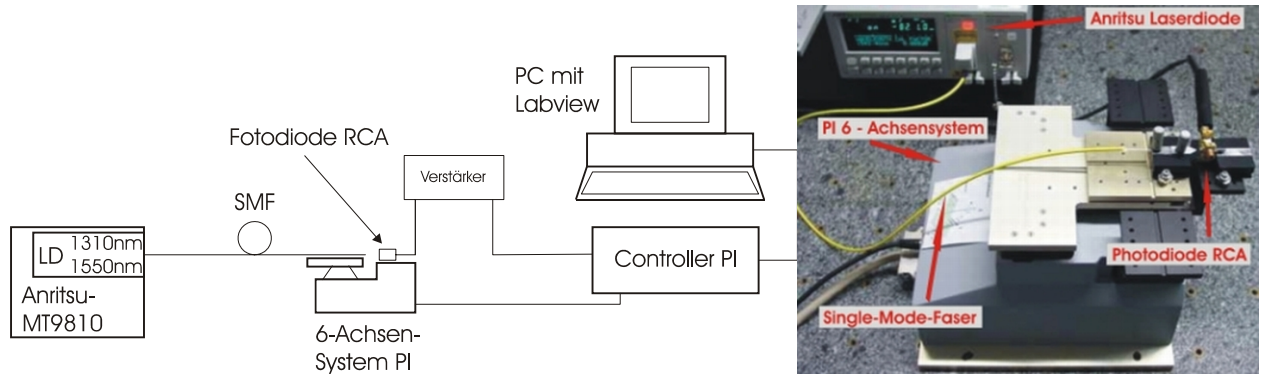

Fig. 7: measurement set-up a) schematic        b) photograph of set-up

As light source the MU951001A integrated in the optical test set MT9810A of Anritsu was used. The module consists of two Fabry Perot laser diodes with the wavelengths 1310nm and 1550nm, which can be alternatively selected. For the detector a photodiode from RCA (InGaAs), with 50µm active surface was used. The fiber and the receiver were installed on a motion controller system of PI, which was controlled by an additional controller unit. However, the controller serves only as interface for the PC where the measurement data recording and calculating is performed.

## 6.  Results

The SMF was measured with different variations by a function of measuring points and incrementations (distance of the measuring points) at the wavelengths of 1550nm and 1310nm. No significant difference can be detected in the result between all variations. Fig. 8 shows the optical mode field diameters determined with the photodiode. We expected a value of 9µm, according to the International Telecommunication Union standard. It is to be recognized that the values measured without distance actor are mostly too small. Values of $2w_0 \approx 7,4µm$ with 1310nm and $2w_0 \approx 8µm$ were determined with 1550nm. This deviation is connected with the distance between DUT and sensor, since this distance could not be determined with the necessary accuracy, without the risk of destruction of the diode.

In order to use this new measuring method it seems to be necessarily to calibrate the entire measuring system on the distance from the test object to the photo receiver since a pre-experimental determining of this distance is not easy to perform. This determination could be achieved by means of highly exact spacer sensors. However, additional equipment must be reinserted. A clearly simpler approach is to make use of standardized mode field width of a single mode fiber at 1310nm. The standard mode field must be measured at the start of the series of measurements with assistance of a calibration measurement with 1310nm with a single mode fiber. The calibration factor for the distance can easily be determined, which would be used with the further process of the same series of measurements. Calculating the distance calibration factor in our measurement we got the exact mode field widths conforming with the ITU-Standard at λ=1310nm:

$F_K$=1,21945

The gate time for the operation was less than one minute, which represents a substantial improvement in relation to conventional measuring procedures. In order to reduce the absolute gate times per measurement of a test object further, the employment of a CCD camera is conceivable. This would have the advantage that the measurement of the mode field can be performed in only one step and gate times will fall down below one second.

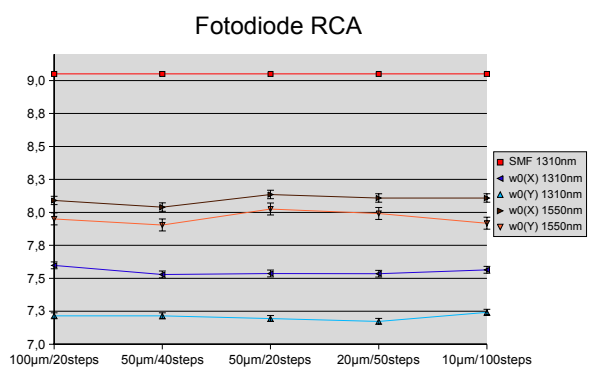

Fig. 8: Measured mode field widths with different detectors and wavelengths without distance calibration

## 7. Summary

We presented a new method for spotsize-measurement for singlemode optical components. In comparison to the classical far field-method and near filed method we had shown that the new method is ideally for use in automated systems e.g. at the production of lasers, where the lasers must be characterized very quickly on bar. The measurement time can be shortened to parts of a second in combination with additionally P-I-characterizations.

In combination with a Labview program the planar measurement-method shows great potential for automated high-speed high precise mode field measurement in accordance with ITU Recommendation G.652.

## 8. Acknowledgement

The Department of Research and Development of the Federal Republic of Germany and the Ministry of Sachsen-Anhalt supported this work. We want to thank M. Biletzke from the Heinrich-Hertz-Institute for delivering the DFB laser diodes for the testing of the field parameters.